\documentclass[preprint,amsmath,amssymb,aps,pre]{revtex4-1} 
\pdfoutput=1
\usepackage[normalem]{ulem}
\usepackage{graphicx,cancel}
\usepackage{dcolumn}
\usepackage{bm}
\usepackage{color}
\definecolor{oliveGreen}{rgb}{0,0.5,0}
\definecolor{magneta}{rgb}{1,0,1}
\usepackage{amssymb,amsfonts,amsmath}
\usepackage{graphicx}
\usepackage{dcolumn}
\usepackage{bm}
\usepackage{color}
\definecolor{oliveGreen}{rgb}{0,0.5,0}
\definecolor{magneta}{rgb}{1,0,1}
\newcommand{\revII}{}
\newcommand{\rev}{} 
\newcommand{\kpc}{}

\begin{document}
\title{Dependence of the dielectric constant of electrolyte solutions on ionic concentration - a microfield approach}

\author{Nir Gavish} 
\affiliation{Technion -- Israel Institute of Technology, Haifa, Israel}%
\author{Keith Promislow}%
\affiliation{Michigan state University, East Lansing, MI, USA}%

\begin{description}
\item[PACS numbers]{61.20.Qg,77.22.Gm,77.22.Ch}
\end{description}
\begin{abstract}
{We present a novel microfield approach for studying the dependence of the orientational polarization of the water in aqueous electrolyte solutions upon the salt concentration and temperature.  The model takes into account the orientation of the solvent dipoles due to the electric field created by ions, and the effect of thermal fluctuations.   The model predicts a dielectric functional dependence of the form $\varepsilon(c)=\varepsilon_w-\beta L(3\alpha c/\beta),\quad\beta=\varepsilon_w-\varepsilon_{\rm ms}$, where~$L$ is the Langevin function,~$c$ is the salt concentration,~$\varepsilon_w$ is the dielectric of pure water,~$\varepsilon_{\rm ms}$ is the dielectric of the electrolyte solution at the molten salt limit, 
and~$\alpha$ is the total excess polarization of the ions. The functional form gives a remarkably accurate description of the dielectric constant for a variety of salts and a wide range of concentrations.}
\end{abstract}

\maketitle
\section{Introduction}
It is difficult to overstate the importance of aqueous electrolyte solutions in biological and electrochemical systems.  There has been extensive study of physical properties of such solutions over the last 120 years.  
The century-old Poisson-Boltzmann (PB) theory gives a simple and powerful description of {\kpc low molarity solutions}, taking into account only Coulombic interactions on a mean-field level, while treating the aqueous solution as a continuous and homogeneous dielectric medium with a dielectric constant~$\varepsilon_s$.  The dielectric constant, however, is typically heterogenous and depends, among other factors, on the local concentration of ions.  Heterogeneity of the dielectric constant significantly influences the structure of the electric double layer region~\cite{ruckenstein2003specific,hatlo2012electric,gavish2014structure,ben2011dielectric,nakayama2015differential,li2015mean}, and effects electrokinetic phenomena, including electro-osmosis and electrophoresis~\cite{zhao2013influence}, as well as charge transfer~\cite{joung2015effect}.  {\revII 
Furthermore, a good understanding of the dielectric properties of a solvent is essential for an accurate description of molecular-level studies of macro biomolecules. For example, some models of globular protein solutions employing a continuum solvent model arbitrarily ascribe a uniformly dielectric value, typically to a value which reflects a pure water solution~\cite{abramo2011molecular,pellicane2013theoretical}. In reality, however, the solvent phase typically contains both an electrolyte and a pH buffer which change the overall ionic strength, yielding a dielectric layer surrounding the immersed macromolecule. The incorporation of variable ionic densities near a macromolecule, and their impact on the local dielectric value, are an important enhancement  of implicit-solvent models~\cite{chen2014modeling,li2014ionic}.}

The first systematic experimental study of the dielectric properties of salt-water solutions was conducted in 1948 by Hasted et al.~\cite{Hasted1948}.  In this work, the static dielectric constant of a solution was observed to decrease with the salt concentration, a phenomena called~{\em dielectric decrement}.  
Intuitively, the dielectric decrement stems from the fact that the local electric field generated by each ion inhibits
the external applied field.  The polar water molecules tend to align with the local ionic field, creating a {\em hydration shell} around the ion, lowering the response of the water molecules to the external field and hence lowering the dielectric constant. In dilute solutions (typically salt concentrations less than 1.5M) the dielectric decrement is linear, 
\begin{equation}\label{eq:linearDecrement}
\varepsilon=\varepsilon_w-\alpha\, c,
\end{equation}
where~$\varepsilon_w$ is the dielectric constant of pure water,~$c$ is the salt concentration, and~$\alpha$ is a phenomenological ion-specific parameter, known as the total {\em excess polarization} of the ionic species.  At higher salt concentrations significant deviations from linearity are observed and the dielectric decrement is observed to saturate~\cite{Hasted1948}.

Haggis et al.~\cite{Haggis1952} modelled the observed linear dielectric decrement by considering the hydration shells as small spherical regions with a low dielectric constant, immersed in pure water medium with a high dielectric constant.   The macroscopic dielectric constant of the solution was then computed by homogenization.   The model was later refined by considering the variation of the local dielectric constant near the ions~\cite{Glueckauf1964} and finite-size effects~\cite{Liszi1988}.  The treatment of hydration shells as spheres, however, is justifiable only for dilute solutions (typically less than 1M) for which the hydration shells do not overlap.  

To go beyond dilute solutions, Levy, Andelman and Orland~\cite{Andelman2012} used a field-theory approach to calculate the average dielectric constant around each ion at the mean-field level, and accounted for hydration shell overlap via a one-loop correction.  The resulting prediction of the dielectric constant affords a good fit to data from a large range of concentrations of different salts using a single fit parameter related to the effective size of the ions. 

In this work, we develop a model for the dielectric response of water molecules in electrolytes with high salt concentrations (in excess of 1.5M) in which the solvation shells of the ions strongly overlap.  The model is not based upon the field around a single ion.  Rather, it assumes the water dipoles are influenced by an aggregate of ions:  the local microfield acting on a water dipole arises  from a surrounding configuration of ions.  {\revII Linking the acquired analytic prediction for the dielectric constant at the high concentration regime to the known theory for the dilute case, yields} an analytic prediction for the solution dielectric constant as a function of salt concentration~$c$,
\begin{equation}\label{eq:thePrediction_eps}
\varepsilon(c)=\varepsilon_w-\beta L\left(\frac{3\alpha}{\beta} c\right),\qquad \beta=\varepsilon_w-\varepsilon_{\rm ms},
\end{equation}
where~$L$ is the Langevin function
\begin{equation*}
L(v)=\coth(v)-\frac1v,
\end{equation*}$\varepsilon_w$ is the dielectric of the pure solvent,~$\varepsilon_{\rm ms}$ is the limiting dielectric constant of the highly concentrated electrolyte solution, a.k.a. the molten salt dielectric, and~$\alpha$ is the total excess polarization of the ions.  This functional form gives a remarkably accurate prediction of the static dielectric constant over a large range of concentrations of 1:1 salts using only a single fitting parameter~$\varepsilon_{\rm ms}$ \footnote{We distinguish between fit parameters, e.g., ‘effective ion size’, which are model-dependent and empirical parameters such as ‘crystal ion radius’ which are model-independent and can be measured experimentally. Accordingly, the parameter~$\varepsilon_{\rm ms}$ is effectively regarded as a fit parameter.}.

The static dielectric constant {\kpc in the high concentration regime} takes the functional form
\begin{equation}\label{eq:eps_orient}
\varepsilon(c)=\varepsilon_{\rm ms}+\frac1{\varepsilon_0}\left.\frac{\partial {P_{\rm water}}(c,E_{\rm ex})}{\partial E_{\rm ex}}\right|_{E_{\rm ex}=0},
\end{equation} 
where~$E_{\rm ex}$ is external field intensity,~$P_{\rm water}$ is the orientational polarization due to orientation of water dipoles in the direction of applied field,~$\varepsilon_0$ is the vacuum permittivity, and~$\varepsilon_{\rm ms}$ is the contribution to the dielectric arising from molecular polarization and orientational polarization of ion-pairs.  
The aim of this work is to derive a mechanism for the dependence of the orientational polarization~$P_{\rm water}$ upon~$c$ via the dependence 
of the ionic microfield upon salt concentration. 

\section{The ionic microfield} We consider a system of 1:1 ions which is globally charge-neutral, and aim to find the distribution of the intensity of electric field created by the ions, the ionic microfield.   Such a distribution is influenced by correlations between anions and cations.  Roughly speaking, each ion is surrounded by an oppositely charged ``ionic atmosphere''.  
{\kpc An expression for the probability density function~$f(E_{\rm ion};c)$ of the ionic field intensity  was derived by Rozental~\cite{EionStatistics77} by treating each ion and its oppositely charged ``ionic atmosphere'' as a dipole, and analyzing the microfield statistics of the dipole configuration to derive the probability density function}
\begin{subequations}\label{eq:pdfEion}
\begin{equation} \label{eq:pdfEiona}
f(E_{\rm ion};c)=\frac{4}{\pi}\frac1{E_{\rm ion}^*}\frac{(E_{\rm ion}/E_{\rm ion}^*)^2}{\left[1+(E_{\rm ion}/E_{\rm ion}^*)^2\right]^2},
\end{equation}
where~$E_{\rm ion}^*$ is the most probable ionic field intensity, satisfying, 
\begin{equation}\label{eq:pdfEionb}
\frac{p}{kT}E_{\rm ion}^*=\alpha^* c.
\end{equation}
\end{subequations}
{\kpc Here~$p$ is the electric dipole moment of water}, and~$\alpha^*$, with units of $M^{-1}$, is proportional to the electric field screening length,  
and serves as a dimensional constant of proportionality between the normalized ionic field intensity and the ionic concentration. 
 Notably, the relation $E_{\rm ion}^*\propto c$ stems from the correlation between the positive and negative ions.  Indeed, the microfield distribution due to uncorrelated positive and negative ions is given by the Holtzmark distribution with~$E_{\rm ion}^*\sim c^{2/3}$~\cite{Holtsmark1919}.

The effect of water molecules on the ionic field intensity is neglected in these calculations, i.e., the ionic field is considered as if the ions were in vacuum.  The justification for this simplification is that at high concentrations, the water/ion ratio does not allow for efficient screening by water molecules.
For example, at 1M concentration, the average distance between ions and their counter-ions can not exceed 1.2nm.  At this separation, at most 3 water molecules can reside along the line segment connecting the ionic centers, hence the screening of electric field by the water is limited.
In contrast, at 1mM solutions, ion separation distances average around 12nm, with 50-100 water molecules engaged in screening.  

A key feature of the model is that $\alpha^*$, which
relates ionic concentrations to screening length, particularly
ionic screening, is independent of ionic concentrations
above 1.5-2M.  At these concentrations inter-ionic distances relate weakly
to the concentration, and the ability of ions
to redistribute to improve effective screening becomes limited due to finite size effects and thermal fluctuations.  As a result, the effective screening length saturates. 

\section{Orientational polarization~$P_{\rm water}$ due to water dipoles}
We first determine the contribution of a single water dipole to the orientational (dipolar) polarization by introducing an ionic field to the standard Langevin dipole analysis, see~\cite[sect. 4.6]{Jackson} or~\cite[p. 214]{Matveyev} for details.
Consider a water dipole~${\bf p}$ surrounded by point-like ions under an applied external field~${\bf E}_{\rm ex}$.
The electric field due to the ions, in the absence of water dipoles, is denoted by~${\bf E}_{\rm ion}$.  
The potential energy~$W$ of the dipole~${\bf p}$ is given by
\begin{equation*}
W=-{\bf p}\cdot ({\bf E}_{\rm ion}+{\bf E}_{\rm ex}).
\end{equation*}
The energy $W$ is minimized when the dipole is aligned with the field~${\bf E}_{\rm ion}+{\bf E}_{\rm ex}$, hence the dipole orientation is a trade-off between the tendency of the dipole to align with the field~${\bf E}_{\rm ion}+{\bf E}_{\rm ex}$ and thermal fluctuations that disrupt this ordering.
The contribution of a given dipole~${\bf p}$ to the polarization due to the external field~${\bf E}_{\rm ex}$ is~$\mbox{Proj}_{{\bf E}_{\rm ex}}({\bf p}-{\bf p}_0)$
where~${\bf p}_0$ is the average orientation of the dipole~${\bf p}$ in the absence of an external field, i.e., in the direction of~${\bf E}_{\rm ion}$,
\begin{equation*}
{\bf p}_0=\frac{{\bf E}_{\rm ion}}{|{\bf E}_{\rm ion}|}|{\bf p}|.
\end{equation*}
The expectation of the contribution of the dipole~${\bf p}$ to the external field~${\bf E}_{\rm ex}$ is given by the Boltzmann average
\begin{equation}\label{eq:BoltzmannAvg}
P^{\rm local}_{\rm water}=\frac{\int e^{-\frac{W}{kT}} \mbox{Proj}_{{\bf E}_{\rm ex}}({\bf p}-{\bf p}_0)}{\int e^{-\frac{W}{kT}}},\qquad \end{equation}
where~$E_{\rm ion}=|{\bf E}_{\rm ion}|,$ and~$E_{\rm ex}=|{\bf E}_{\rm ex}|$.
This integral can be explicitly evaluated~\cite{Nielsen07}, 
\begin{equation}\label{eq:Plocal}
P^{\rm local}_{\rm water}=pL\left(\frac{pE_{\rm ex}}{kT}\right)\left[1-L\left(\frac{pE_{\rm ion}}{kT}\right)\right],\qquad \end{equation}
where~$p=|{\bf p}|$, see Appendix~\ref{app:BoltzmannAvg} for details. 

Expression~\eqref{eq:Plocal} quantifies the contribution of a single water dipole {\kpc influenced by an ionic field with intensity~$E_{\rm ion}$} to the orientational polarization.  The second stage of the derivation considers the aggregate contribution~$P_{\rm water}$ of all water dipoles to the orientational polarization {\kpc by considering the statistics of the ionic field intensity {\em experienced} by the water dipoles.  Under the assumption that the water dipoles are uniformly distributed in space, the local ionic field intensity influencing the water dipoles is distributed according to~$f(E_{\rm ion};c)$.  Deviations from uniform distribution, however, do occur.  { \kpc Indeed, there is a minimal water dipole - ion separation, and this effect serves to exclude water dipoles from regions} where the microfield intensity is the highest.  To account for this phenomena, we assume there is a bound~$\alpha_ME_{\rm ion}^*$ on the ionic field intensity felt by water molecules, {\rev which is proportional to the most probable ionic field intensity~$E_{\rm ion}^*$.}  Incorporating this limit,}~$P_{\rm water}$ is given by
\begin{equation}\label{eq:aggregatedP}
P_{\rm water}(c)=N_w\int_0^{\alpha_M E_{\rm ion}^*} f(E_{\rm ion};c)P^{\rm local}_{\rm water}(E_{\rm ion}) dE_{\rm ion},
\end{equation}
where~$P^{\rm local}_{\rm water}$ is given by~\eqref{eq:Plocal},~$f(E_{\rm ion};c)$ is given by~\eqref{eq:pdfEion}, and~$N_w$ is the number of water molecules per unit volume.
The Langevin function can be approximated by
\begin{equation}\label{eq:LangevinLargev}
L(v)\approx 1-1/v
\end{equation}
for~$v>v_L=3$.  
Using{\rev~\eqref{eq:pdfEionb}} and~\eqref{eq:LangevinLargev}  to simplify~\eqref{eq:Plocal} in the integral~\eqref{eq:aggregatedP}, see Appendix~\ref{app:Pwater_expansion} for details, yields 
\begin{equation}\label{eq:pdf_to_L_tail}
{P}_{\rm water}(c)\approx pL\left(\frac{pE_{\rm ex}}{kT}\right)\frac{\mu}{\alpha^* c},\qquad \mu=\frac{2N_w\alpha_M^2\,}{\pi(1+\alpha_M^2)}.
\end{equation}
 Expression~\eqref{eq:pdf_to_L_tail} quantifies the aggregate contribution of all water dipoles to the orientational polarization~$P_{\rm water}$.  As expected,~$P_{\rm water}\to0$ as~$c\to\infty$ (or~$E_{\rm ion}\to \infty$), since the water dipoles become `immobilized' by the strong ionic field.
 Substituting~\eqref{eq:pdf_to_L_tail} into~\eqref{eq:eps_orient} gives the functional form 
{\kpc\begin{equation}\label{eq:eps_tail}
\varepsilon(c)=\varepsilon_{\rm ms}+\frac{p^2}{3kT\varepsilon_0}\frac{\mu}{\alpha^*c}.
\end{equation}}
Relation~\eqref{eq:eps_tail} is derived in the high concentration regime~$\alpha^*c\gg1$, in which the electric field screening length, $\alpha^*$,
may be taken independent of ionic concentration.  For more dilute regimes the electric field screening length agrees with
the Debye length and scales like $1/\sqrt{c}$.  Although the primary goal of this derivation is to obtain a prediction for~$\varepsilon(c)$ in the high concentration regime, {\kpc there is utility in developing an approximation that is compatible with known results in the dilute regime.  {\revII To do so, we develop a composite expansion }
\begin{subequations}\label{eq:pdf_to_L}
\begin{equation}\label{eq:pdf_to_L_relation}
\varepsilon(c)=\varepsilon_w-\beta L\left(\frac{3\alpha}{\beta}c\right),
\end{equation}
{\revII where the two parameters~$\alpha$ and~$\beta$ are determined by a matching condition with the high concentration limit of~\eqref{eq:pdf_to_L_relation}}
\begin{equation*}
\varepsilon(c)=\varepsilon_w-\beta L\left(\frac{3\alpha}{\beta}c\right)\approx \varepsilon_w -\beta+\frac{\beta^2}{3\alpha c},\quad c\gg1,
\end{equation*}
yielding the relation
\begin{equation}\label{eq:pdf_to_L_parms}
\beta=\varepsilon_w-\varepsilon_{\rm ms},\quad \alpha^*=\frac{2\,\mu\, p^2}{\pi k_BT\varepsilon_0}\frac{\alpha}{\beta^2}.
\end{equation}
\end{subequations}
{\revII The composite expansion~\eqref{eq:pdf_to_L} is compatible with the prediction~\eqref{eq:eps_tail} for~$\varepsilon(c)$ in the high concentration regime, see~\eqref{eq:LangevinLargev}, as well as with the prediction~\eqref{eq:linearDecrement} in the dilute regime.}
In the sequel, we show that the composite expansion~\eqref{eq:pdf_to_L}  yields a uniformly
valid and highly accurate approximation of~$\varepsilon(c)$.  An artifact of the composite expansion~\eqref{eq:pdf_to_L}, however, is that the excess polarization parameter~$\alpha$, which is related to ion-water interactions, is described in terms of parameters related to ion-ion interactions.  Namely,~$\alpha=\alpha(\alpha^*,\varepsilon_{ms})$, see~\eqref{eq:pdf_to_L_parms}.}

\section{Agreement with experimental data} 
\begin{figure*}[ht!]
\scalebox{1}{\includegraphics{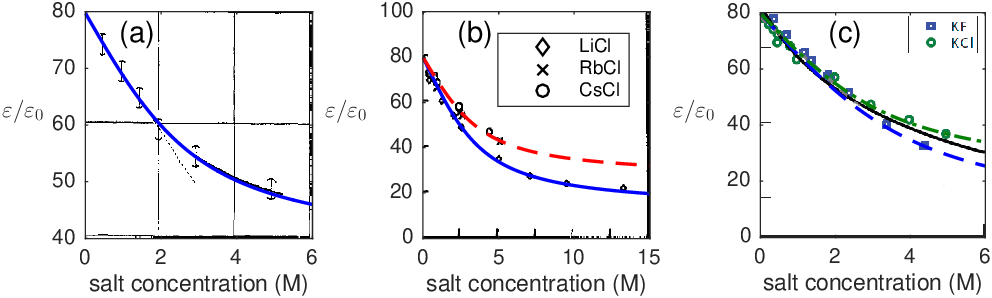}}
\caption{(color online) Comparison of the predicted dielectric constant~\eqref{eq:thePrediction_eps}, with experimental data as function of ionic concentration~$c$ for various salts. Here~$\alpha$ is extracted directly {\rev from the slope of~$\varepsilon(c)$ at~$c\ll1$ for each experimental dataset}, and~$\varepsilon_{\rm ms}$ is fitted separately for each type of electrolyte. 
a: Data for NaCl salt from~\cite{Hasted1948}, compared to~\eqref{eq:thePrediction_eps} with~$\alpha=11.5$ and~$\varepsilon_{\rm ms}=51.42$.
b: Data from~\cite{Wei92}, where the parameters for RbCl and CsCl salts ({\color{red} $--$}) are~$\alpha=11$ and~$\varepsilon_{\rm ms}=26.48$ and for LiCl ({\color{blue}$-$}) the parameters are~$\alpha=14$ and~$\varepsilon_{\rm ms}=12.5$.  
c: Figure 2(b) from~\cite{Andelman2012} where parameters are~$\alpha=15$ and~$\varepsilon_{\rm ms}=4.7$ for KF ({\color{blue} $--$}) and~$\alpha=14$ and~$\varepsilon_{\rm ms}=19.70$ for KCl ({\color{oliveGreen} $.-$}).  Solid black curve is the prediction obtained using the field-theory approach~\cite{Andelman2012}.  \label{fig:dielectricDeptFit} }
\end{figure*}

The functional relation~\eqref{eq:thePrediction_eps} is validated against six different sets of experimental data, where~$\varepsilon_{\rm ms}$ is fitted separately for type of electrolyte, and~$\alpha$ is extracted directly from the data from the slope of~$\varepsilon(c)$ \footnote{The extracted values are in agreement with the literature values for~$\alpha$ in the cases they are available~\cite{Hasted1948}, except the literature value for KF~\cite{Hasted1948} ($\alpha=10\pm2$) which does not describe later experimental measurements presented in~\cite{Wei92}.}.
Figure~\ref{fig:dielectricDeptFit}a presents the experimental data of Hasted et al.~\cite{Hasted1948} for an~NaCl solution at 21$^\circ$.  
Prediction~\eqref{eq:thePrediction_eps} agrees very well with the experimental data over the full range of~$0\le c\le 6M$.     
Figure~\ref{fig:dielectricDeptFit}b presents the experimental data of Wei et al.~\cite{Wei92} for LiCl, RbCl and CsCl solutions at 25$^\circ$.  The predicted dielectric~\eqref{eq:thePrediction_eps} remains accurate for LiCl data at concentrations as high as 13M.
Finally, Figure~\ref{fig:dielectricDeptFit}c shows data taken from~\cite{Wei92}, which were compiled and presented in~\cite{Andelman2012}, together with the prediction obtained using the field-theory approach~\cite{Andelman2012} (solid black).  

\begin{figure}[ht!]
\scalebox{1}{\includegraphics{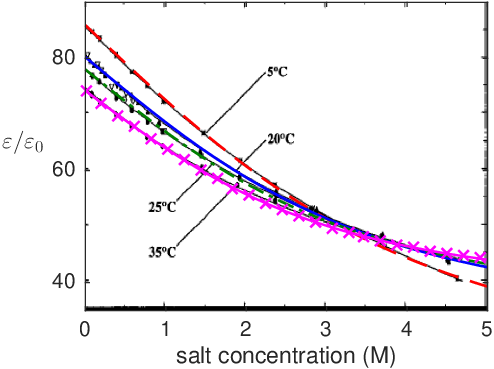}}
\caption{
(color online) Comparison of the predicted dielectric constant~\eqref{eq:thePrediction_eps}, with experimental data from~\cite{Buchner99} as a function of ionic concentration~$c$ for NaCl at various temperatures. 
Data for~$T=5^\circ$C fit with~$\alpha=13.7$ and~$\varepsilon_{\rm ms}=15.67$ ({\color{red} {\bf -\,-}}), for~$T=20^\circ$C with~$\alpha=12$ and~$\varepsilon_{\rm ms}=27.27$ ({\color{blue} {\bf --}}), for~$T=25^\circ$C with~$\alpha=11.5$ and~$\varepsilon_{\rm ms}=30.08$ ({\color{oliveGreen} {\bf .-}}) and for~$T=35^\circ$C with~$\alpha=10.7$ and~$\varepsilon_{\rm ms}=34.07$ ({\color{magenta} {\bf x-}}).
\label{fig:NaCl_temp} } \end{figure}

The model readily incorporates the dependence of the dielectric constant upon temperature~$T$ through functional relations of the parameters $\alpha$ and $\varepsilon_{\rm ms}$ upon temperature.  The functional relation~\eqref{eq:thePrediction_eps} is validated against four different sets of experimental data, obtained from Buchner et al.~\cite{Buchner99}, for~NaCl at various temperatures. Here~$\varepsilon_{\rm ms}$ is fitted separately for each temperature, and~$\alpha$ is extracted directly from each experimental dataset, see Figure~\ref{fig:NaCl_temp}. 

\section{Conclusions} We have presented a model which gives a remarkably accurate fit for the static dielectric constant of aqueous electrolyte solutions.  Modelling the contribution of orientational polarization of water molecules to the dielectric constant requires a description of the ionic field effecting the water molecules in the solution.
In contrast to the classic approach which assumes that the ionic field at a point is due to a dominant ion, the model derivation uses statistics of the ionic microfield to 
characterize the ionic configuration affecting the water molecules in the solution. This approach is naturally suited for  concentrated solutions for which the local electric 
field arises from a configuration of several ions.  Microfield statistics are widely used to describe strongly coupled Coulomb systems in Plasma physics, see~\cite{demura2009physical} and references within, as well as in Astronomy~\cite{Holtsmark1919} and Astrophysics~\cite{chandrasekhar1943statistics1,chandrasekhar1943statistics2}.  
To the best of our knowledge, however, this is the first work to utilize a microfield approach in electrolyte solutions.

Electrolyte solutions are highly complex mixtures, and the model neglects numerous structural elements: water/water interactions, water/ion-pair interactions, finite size effects, reaction field effects, influence of the hydrogen bond network, and the decrease in water molarity as ionic concentration increases.  While a systematic study of these effects is clearly important, it is plausible that they will contribute perturbatively.    In addition, the model accounts for possible contribution to the static dielectric constant due to orientational polarization of ion-pairs via a fitting parameter~$\varepsilon_{\rm ms}$, which corresponds to the limiting dielectric constant of highly concentrated electrolyte solutions.  {\revII It is interesting to note that the fitted parameter~$\varepsilon_{\rm ms}$ does not seem to follow a clear trend, e.g., within a series of alkali chloride.} 
Recent studies have focused on development of a theory for the value of~$\varepsilon_{\rm ms}$ in ionic liquids~\cite{lee2014room,zarubin2015static}.    
Extension of such a theory to concentrated electrolyte solutions, and providing and recovery of the dependence of ionic orientational polarization upon ionic concentration is the subject of future work.

\subsection*{Acknowledgment}
The first auhtors acknowledges support from the Technion VPR fund and from EU Marie--Curie CIG grant 2018620, while the second author acknowledges support from the US National Science Foundation through DMS-1409940.
\appendix

\section{Explicit evaluation of the integral for~\eqref{eq:BoltzmannAvg}.}\label{app:BoltzmannAvg}
Using a spherical coordinate system~$(r,\theta,\phi)$ for the three vectors
\[
{\bf E}_{\rm ex}=(E_{\rm ex},0,0),\quad {\bf E}_{\rm ion}=(E_{\rm ion},\theta_{\rm ion},\phi_{\rm ion}),\quad 
{\bf p}=(p,\theta,\phi),
\]
the integral~\eqref{eq:BoltzmannAvg} takes the form
\begin{subequations}\label{eq:originalIntegral}
\begin{equation}
P_{\rm local}=p\frac{\int e^{-\frac{W}{kT}} \left(\cos\theta-\cos\theta_{\rm ion}\right)\sin\theta\sin\theta_{\rm ion}d\theta d\theta_{\rm ion}d\phi d\phi_{\rm ion}}{\int e^{-\frac{W}{kT}}\sin\theta\sin\theta_{\rm ion}d\theta d\theta_{\rm ion}d\phi d\phi_{\rm ion}},
\end{equation}
where
\begin{equation}
W=-pE_{\rm ion}\left[\cos\theta_{\rm ion}\cos\theta+\sin\theta_{\rm ion}\sin\theta\cos(\phi-\phi_{\rm ion})\right]-pE_{\rm ex}\cos\theta.
\end{equation}
\end{subequations}

\begin{figure}[ht!]
\scalebox{.5}{\includegraphics{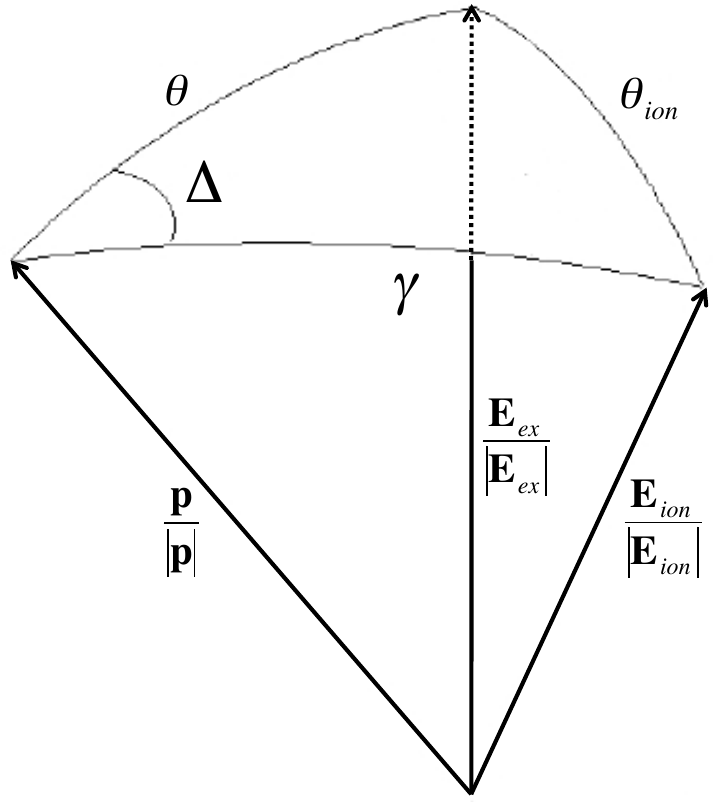}}
\caption{The spherical triangle created by the unit vectors in the directions of~${\bf E}_{\rm ion}$,~${\bf E}_{\rm ex}$ and~${\bf p}$.  \label{fig:vectorDiagram} }
\end{figure}

This integral is evaluated by making the following change of variables
\[(\theta_{\rm ion},\theta,\phi-\phi_{\rm ion})\to(\gamma,\theta,\Delta)\] where~$\gamma$ is the angle between~${\bf E}_{\rm ion}$ and~${\bf p}$, and~$\Delta$ is the vertex angle opposed to~$\theta_{\rm ion}$ in the spherical triangle created by the unit vectors in the directions of~${\bf E}_{\rm ion}$,~${\bf E}_{\rm ex}$ and~${\bf p}$~\cite{Nielsen07}, see Figure~\ref{fig:vectorDiagram}.

The new variables satisfy 
\begin{align}
&\label{eq:COV_cosg}\cos\gamma=\cos\theta\cos\theta_{\rm ion}+\sin\theta\sin\theta_{\rm ion}\cos(\phi-\phi_{\rm ion}),\\
&\label{eq:COV_cosD}\cos\theta_{\rm ion}=\cos\theta\cos\gamma+\sin\theta\sin\gamma\cos\Delta,
\end{align}
where identity~\eqref{eq:COV_cosD} follows from the law of cosines for spherical triangles.

The integral~\eqref{eq:originalIntegral}, therefore, reduces to
\begin{subequations}\label{eq:integralWithGamma}
\begin{equation}
p\frac{\int_{\Delta=0}^{2\pi}\int_{\gamma=0}^{2\pi}\int_{\theta=0}^{\pi}e^{-\frac{W}{kT}}\left(\cos\theta-\cos\theta\cos\gamma-\sin\theta\sin\gamma\cos\Delta\right) J(\theta,\gamma,\Delta) \sin\theta\sin\theta_{\rm ion}d\theta d\gamma d\Delta}{\int_{\Delta=0}^{2\pi}\int_{\gamma=0}^{2\pi}\int_{\theta=0}^{\pi}e^{-\frac{W}{kT}}J(\theta,\gamma,\Delta)\sin\theta\sin\theta_{\rm ion} d\theta d\gamma d\Delta}.
\end{equation}
where
\begin{equation}
e^{-\frac{W}{kT}}=\exp\left[\frac{p}{kT}\left(E_{\rm ion}\cos\gamma+E_{\rm ex}\cos\theta\right)\right].
\end{equation}
\end{subequations}
The Jacobian~$J(\theta,\gamma,\Delta)$ is computed as follows:
Direct differentiation of~\eqref{eq:COV_cosg} yields
\[\begin{split}
&\frac{\partial\phi}{\partial\gamma}=\frac{\sin\gamma}{\sin\theta\sin\theta_{\rm ion}\sin\phi}+\frac{\sin\theta\cos\theta_{\rm ion}\cos\phi-\cos\theta\sin\theta_{\rm ion}}{\sin\theta\sin\theta_{\rm ion}\sin\phi}\frac{\partial\theta_{\rm ion}}{\partial\gamma},\\
&\frac{\partial\phi}{\partial\Delta}=\frac{\sin\theta\cos\theta_{\rm ion}\cos\phi-\cos\theta\sin\theta_{\rm ion}}{\sin\theta\sin\theta_{\rm ion}\sin\phi}\frac{\partial\theta_{\rm ion}}{\partial\Delta}.
\end{split}
\]
Direct differentiation of~\eqref{eq:COV_cosD} yields
\[\begin{split}
&\frac{\partial\theta_{\rm ion}}{\partial\gamma}=\frac{\cos\theta\sin\gamma-\sin\theta\cos\gamma\cos\Delta}{\sin\theta_{\rm ion}},\\
&\frac{\partial\theta_{\rm ion}}{\partial\Delta}=\frac{\sin\theta\sin\gamma\sin\Delta}{\sin\theta_{\rm ion}}.
\end{split}
\]
Thus,
\[
\begin{split}
J(\theta,\gamma,\Delta):=&\frac{\partial\phi}{\partial\gamma}\frac{\partial\theta_{\rm ion}}{\partial\Delta}-\frac{\partial\phi}{\partial\Delta}\frac{\partial\theta_{\rm ion}}{\partial\gamma}=\frac{\sin\gamma}{\sin\theta\sin\theta_{\rm ion}\sin\phi}\frac{\partial\theta_{\rm ion}}{\partial\Delta}\\=&\frac{\sin\gamma}{\sin\theta\sin\theta_{\rm ion}\sin\phi}\frac{\sin\theta\sin\gamma\sin\Delta}{\sin\theta_{\rm ion}}=\frac{\sin^2\gamma}{\sin^2\theta_{\rm ion}}\frac{\sin\Delta}{\sin \phi}.
\end{split}
\]
By the law of sines for spherical triangles,
\[
\frac{\sin\theta_{\rm ion}}{\sin \Delta}=\frac{\sin\gamma}{\sin\phi},
\]
Hence, the Jacobian further reduces to 
\[
J(\theta,\gamma,\Delta)=\frac{\sin\gamma}{\sin\theta_{\rm ion}}.
\]
Substitution of the Jacobian in the integral~\eqref{eq:integralWithGamma} yields
\[
\begin{split}
&P_{\rm local}=\\&\,\,p\frac{\int_{\Delta=0}^{2\pi}\int_{\gamma=0}^{2\pi}\int_{\theta=0}^{\pi}\exp\left[\frac{p}{kT}\left(E_{\rm ion}\cos\gamma+E_{\rm ex}\cos\theta\right)\right]\left(\cos\theta-\cos\theta\cos\gamma-\cancel{\sin\theta\sin\gamma\cos\Delta}\right)\, \sin\theta\sin\gamma d\theta d\gamma d\Delta}{\int_{\Delta=0}^{2\pi}\int_{\gamma=0}^{2\pi}\int_{\theta=0}^{\pi}\exp\left[\frac{p}{kT}\left(E_{\rm ion}\cos\gamma+E_{\rm ex}\cos\theta\right)\right]\sin\theta\sin\gamma d\theta d\gamma d\Delta}.
\end{split}
\]
This integral can be evaluated directly as is done with the standard evaluation of the integral for the Langevin function, see, e.g.,~\cite[p. 215]{Matveyev} for details. 
{\kpc\section{Asymptotic expansion of~$P_{\rm water}(c)$ at high concentration regime}\label{app:Pwater_expansion}
The expression for~$P_{\rm water}(c)$, see~\eqref{eq:aggregatedP}, reads  as
\[
P_{\rm water}(c)=\mu_c\int_0^{\alpha_ME_{\rm ion}^*}\frac{(E_{\rm ion}/E_{\rm ion}^*)^2}{\left[1+(E_{\rm ion}/E_{\rm ion}^*)^2\right]^2}\left[1-L\left(\frac{pE_{\rm ion}^*}{kT}\frac{E_{\rm ion}}{E_{\rm ion}^*}\right)\right] \frac{dE_{\rm ion}}{E_{\rm ion}^*},\qquad \mu_c=\frac{4N_w}{\pi}L\left(\frac{pE_{\rm ex}}{kT}\right)p.
\]
Making the change of variables~$x=E_{\rm ion}/E_{\rm ion}^*$ under the integral yields
\[
P_{\rm water}(c)=\mu_c\int_0^{\alpha_M} \frac{x^2}{(x^2+1)^2}\left[1-L(\alpha^*c x)\right] \,dx,
\]
To evaluate the integral, we split it into two regimes, from $0$ to $x^*$ and $x^*$ to $\alpha_M$,
where $x^*=\frac{\gamma}{\alpha^*c}$ with $\gamma$ chosen so that
\[
\frac12|\log\gamma|\ll \gamma\ll\alpha_M.
\]
The choice of~$\gamma$, and hence of~$x^*$, is made so that for any~$x^*\le x\le \alpha_M$, the Langevin function can be asymptotically approximated up to an exponentially small error, 
\[
1-L(\alpha^*c x)= \frac{1}{\alpha^*c x}+O(e^{-2\alpha^*c x}).
\]
Therefore,
\[
\int_{x^*}^{\alpha_M}\frac{x^2}{(x^2+1)^2}\left[1-L(\alpha^*c x)\right] dx=
\int_{x^*}^{\alpha_M}\frac{x^2}{(x^2+1)^2}\left[\frac{1}{\alpha^*c x}+O(e^{-2\alpha^*c x})\right] dx=\frac{\alpha_M^2}{1+\alpha_M^2} \frac{1}{2\alpha^* c}+O\left(\frac{1}{c^3}\right)
\]
On the other end,~$1-L(\alpha^*c x)\le 1$.  Thus,
\[
\int_{0}^{x^*}\frac{x^2}{(x^2+1)^2}\left[1-L(\alpha^*c x)\right] dx \le \int_{0}^{x^*}\frac{x^2}{(x^2+1)^2}dx=\frac{(x^*)^3}{3}+O\left((x^*)^5\right)=O\left(\frac{1}{c^3}\right).
\]
Overall, we find that for~$c\gg1$,
\[
P_{\rm water}(c)=\frac{\alpha_M^2\,\mu_c}{1+\alpha_M^2} \frac{1}{2\alpha^* c}+O\left(\frac{1}{c^3}\right).
\]
}

%

 \end{document}